# Optimized phonon band discretization scheme for efficiently solving the non-gray Boltzmann transport equation

Yue Hu, Yongxing Shen, Hua Bao[*]

University of Michigan–Shanghai Jiao Tong University Joint Institute, Shanghai Jiao Tong University, Shanghai 200240, P. R. China

**Abstract**

Phonon Boltzmann transport equation (BTE) is an important tool for studying the nanoscale thermal transport. Because phonons have a large spread in their properties, the non-gray (i.e. considering different phonon bands) phonon BTE is needed to accurately capture the nanoscale transport phenomena. However, BTE solvers generally require large computational cost. Non-gray modeling imposes significant additional complexity to the numerical simulations, which hinders the modeling of real nanoscale systems. In this work, we address this issue by a systematic investigation on the phonon band discretization scheme using real material properties of four representative materials, including silicon, gallium arsenide, diamond, and lead telluride. We find that the schemes used in previous studies require at least a few tens of bands to ensure the accuracy, which requires large computational costs. We then propose an improved band discretization scheme, in which we divide the mean free path domain into two subdomains, one on either side of the inflection point of the mean free path accumulated thermal conductivity and adopt the Gauss-Legendre quadrature for each subdomain. With this scheme, the solution of phonon BTE converges (error < 1%) with less than 10 phonon bands for all these materials. The proposed scheme allows to significantly reduce the time and memory consumption of the numerical BTE solver, which is an important step towards large-scale phonon BTE simulation for real materials.

## I. Introduction

Due to the miniaturization of semiconductor devices [1] in the past few decades, the characteristic length of thermal transport has been reduced to the nanoscale [2,3]. As the scale is comparable to the phonon mean free path, it is necessary to solve the phonon Boltzmann transport equation (BTE) instead of the heat diffusion equation for the thermal transport in semiconductors [4–6].

---

[*] To whom correspondence should be addressed. Email: hua.bao@sjtu.edu.cn (HB)

Before phonon BTE, there has been a long history of solving BTE for other particles including electrons, photons, neutrons, and fluid particles [4,7–9]. Many numerical methods have been developed, such as the lattice Boltzmann method [10], Monte-Carlo method [11], discrete ordinates method (DOM) [9]. These methods have wide applications including simulating rocket propulsion (photons) [9], nuclear reactions (neutrons) [8], microchannel flows (fluid particles) [12], and electrical performance of electronic devices (electrons) [1]. All these methods have been adopted to solve phonon BTE for thermal transport [13–19]. However, unlike other particles, phonons have a unique feature that there is a large spread in phonon properties due to the phonon dispersion and polarization [20,21]. Most importantly, the mean free path (MFP) of phonons can span for several orders of magnitude (for example, five orders of magnitude for silicon) [22], which is quite different from most of other particles. For example, the electron MFP can usually be assumed to be a constant [23,24]. However, the phonon MFP spans a long range, and the gray model that assumed the constant MFP is invalid [25]. In order to capture such a feature, the non-gray version of numerical solvers for phonons has been developed [26–34]. The non-gray model resolves the large spread in phonon properties by band discretization, which discretizes the phonon dispersion into a set of phonon bands. These bands are solved separately and then coupled through an energy conservation equation. Due to the lack of accurate phonon properties, previous works mainly adopted non-gray model for qualitatively studying nanoscale thermal transport phenomena such as size effect and nonequilibrium temperatures [20,26,27,32,35,36]. These studies adopt theoretical models for phonon properties or only collected dispersion and polarization in one crystal orientation. Moreover, most of the studies take toy problems for demonstration purposes, which do not require large computational cost. Therefore, they simply discretized phonon bands by a large number, usually a few tens to a hundred, without conducting a rigorous analysis on the numerical accuracy and efficiency of the band discretization scheme [26–29]. Recent advancements in first-principles calculations [37–39] and experimental tools [40,41] enable resolving the phonon properties for real materials. Such advancements make it possible to take the accurate phonon properties as inputs of BTE to achieve quantitative modeling of nanoscale thermal transport in real systems. At this stage, a study on the phonon band discretization of real properties is highly desirable.

In this work, we systematically investigate the band discretization of real phonon properties, including the detailed analysis of the numerical accuracy and efficiency of the band discretization scheme. Based on the analysis, we propose and demonstrate a band discretization scheme to accurately model real systems with a small band number (<10) in non-gray phonon

BTE for typical materials. This small band number ensures low computational cost of numerical solvers. This manuscript is organized as follows: in Sec. II, we introduce the non-gray phonon BTE, including the equations, boundary conditions, and band discretization. In Sec. III, the numerical accuracy of the band discretization scheme is analyzed for silicon. We find that the previous scheme requires a large number of bands and propose our own scheme, which only requires a small number of bands for silicon. After that, in Sec. IV, the results of analysis in Sec. III are verified by numerically solving two real problems for silicon. The computational cost of numerical solvers is also checked. In Sec. V, we check the performance of our band discretization scheme when applied to other materials. Finally, we provide a summary and conclusions.

## II.  Non-gray Phonon Boltzmann transport equation

### a) Phonon BTE

The steady-state, non-gray BTE under the relaxation time approximation is commonly expressed in a frequency-dependent expression [27]:

$$\mathbf{v}_{\omega,p} \cdot \nabla e_{\omega,p,\mathbf{s}} = -\frac{e_{\omega,p,\mathbf{s}} - e^0_{\omega,p}}{\tau_{\omega,p}}, \tag{1}$$

where $e$ is the distribution function of phonon energy density, and $e^0$ is the energy density distribution function of the equilibrium state, which follows Bose-Einstein distribution [42]. $\omega, p$ is the frequency and branch index. $\mathbf{s}$ is the unit vector in the direction of group velocity $\mathbf{v}$, which is assumed to be isotropic [27,43]. $\tau$ is the relaxation time. According to the energy conservation [27], the $e^0_{\omega,p}$ is related to the $e_{\omega,p,\mathbf{s}}$ as:

$$e^0_{\omega,p} = \frac{\frac{C_{\omega,p}}{4\pi} \int \sum_p \int_{\omega_{\min}}^{\omega_{\max}} \frac{e_{\omega',p,\mathbf{s}}}{\tau_{\omega',p}} d\omega' d\Omega}{\sum_p \int_{\omega_{\min}}^{\omega_{\max}} \frac{C_{\omega',p}}{\tau_{\omega',p}} d\omega'}, \tag{2}$$

where $C$ is the volumetric heat capacity. The information of $C$, $v$ and $\tau$ can be obtained by first-principles calculations[37]. $\Omega$ is the control angle. The heat flux can be obtained by:

$$\mathbf{q} = \int \sum_p \int_{\omega_{\min}}^{\omega_{\max}} \mathbf{v}_{\omega,p} e_{\omega,p,\mathbf{s}} d\omega d\Omega \tag{3}$$

This expression requires the frequency and branch-dependent phonon properties, including the phonon group velocity, phonon relaxation time, and volumetric heat capacity.

The frequency-dependent expression (Eq. (1), (2), and (3)) can be converted into the MFP-dependent expression (derivations can be found in Ref. [44]) :

$$\mathbf{s} \cdot \nabla T_{\mathsf{L},\mathbf{s}} = -\frac{T_{\mathsf{L},\mathbf{s}} - T^0}{\mathsf{L}},$$

$$T^0 = \frac{\frac{1}{4\pi} \iint \int_{\mathsf{L}_{min}}^{\mathsf{L}_{max}} \frac{dK}{d\mathsf{L}} \frac{1}{\mathsf{L}^2} T_{\mathsf{L},\mathbf{s}} d\mathsf{L} d\mathsf{W}}{\int_{\mathsf{L}_{min}}^{\mathsf{L}_{max}} \frac{dK}{d\mathsf{L}} \frac{1}{\mathsf{L}^2} d\mathsf{L}}, \quad (4)$$

$$\mathbf{q} = \int \mathbf{s} \int_{\mathsf{L}_{min}}^{\mathsf{L}_{max}} \frac{dK}{d\mathsf{L}} \frac{1}{\mathsf{L}^2} T_{\mathsf{L},\mathbf{s}} d\mathsf{L} d\mathsf{W}$$

where $T_{\mathsf{L},\mathbf{s}} = e_{\mathsf{L},\mathbf{s}} / C_{\mathsf{L},\mathbf{s}}$ is the phonon temperature distribution and $T^0$ is the corresponding equilibrium temperature (also known as the lattice temperature [27]). The MFP accumulation for thermal conductivity $K(\mathsf{L})$ represents the thermal conductivity of phonons with MFP smaller than $\mathsf{L}$.

$$K(\mathsf{L}) = -\frac{1}{3} \int_{\mathsf{L}_{min}}^{\mathsf{L}} Cv\mathsf{L}' \left(\frac{d\mathsf{L}'}{d\mathsf{W}}\right) d\mathsf{L}'. \quad (5)$$

The previous study showed that MFP-dependent expression has two main advantages over frequency-dependent expression [44]. First, the MFP-dependent expression only requires the MFP accumulated thermal conductivity. The MFP accumulated thermal conductivity not only can be calculated by $C$, $v$ and $\mathsf{L}$ obtained by first-principle calculations, but also can be directly obtained by MFP spectroscopy [40,41]. Second, when applying the MFP-dependent expression, all parameters and the variable are directly related to a single parameter of MFP. One do not need to resolve the phonon branches and crystalline orientations, which is difficult for materials with complex unit cells [42]. Therefore, in the subsequent discussions, we adopt the MFP-dependent expression.

### b) Boundary conditions

Three boundary conditions were widely adopted in phonon BTE [26,27]: (i) thermalizing boundary condition, (ii) specularly reflecting boundary condition, and (iii) diffusely reflecting boundary condition.

At the thermalizing boundaries, all phonons are emitted from the boundary with a temperature of $T_1$:

$$T_{\mathsf{L},\mathbf{s}} = T_1. \quad (6)$$

Specular reflecting boundary condition is expressed as:

$$T_{\Lambda,\mathbf{s}} = T_{\Lambda,\mathbf{s}_r}, \tag{7}$$

where $\mathbf{s}_r$ is the original direction before specularly reflecting to direction $\mathbf{s}$.

Diffusely reflecting boundary condition is expressed as:

$$T_{\Lambda,\mathbf{s}} = \frac{1}{\pi}\int_{\mathbf{s}'\cdot\mathbf{n}>0} T_{\Lambda,\mathbf{s}'}\mathbf{s}'\cdot\mathbf{n}\,d\Omega, \tag{8}$$

where $\mathbf{n}$ is the exterior normal vector of the boundary.

c) **Phonon band discretization**

To numerically solve the non-gray BTE (Eq. (4)), one first needs band discretization. The band discretization transfers the integration associated with the MFP domain into a summation according to a quadrature rule:

$$\int_{\Lambda_{\min}}^{\Lambda_{\max}} \frac{dK}{d\Lambda}\frac{1}{\Lambda^m} T_{\Lambda,\mathbf{s}_a,\nu_i}\,d\Lambda = \sum_{\Lambda_b} w_b \frac{dK}{d\Lambda}(\Lambda_b)\frac{1}{\Lambda_b^m} T_{\Lambda_b,\mathbf{s}_a,\nu_i}, m=1,2 \tag{9}$$

where $w_b$ is the weight of phonon band with MFP $\Lambda_b$ in the quadrature rule. Plugging Eq. (9) into Eq. (4), one can obtain:

$$\mathbf{s}\cdot\nabla T_{\Lambda_b,\mathbf{s}} = -\frac{T_{\Lambda_b,\mathbf{s}} - T^0}{\Lambda_b},$$

$$T^0 = \frac{\frac{1}{4\pi}\int \sum_{\Lambda_b} w_b \frac{dK}{d\Lambda}(\Lambda_b)\frac{1}{\Lambda_b^2} T_{\Lambda_b,\mathbf{s}}\,d\Omega}{g}, \tag{10}$$

$$\mathbf{q} = \int \mathbf{s} \sum_{\Lambda_b} w_b \frac{dK}{d\Lambda}(\Lambda_b)\frac{1}{\Lambda_b} T_{\Lambda_b}\,d\Omega,$$

where $g = \int_{\Lambda_{\min}}^{\Lambda_{\max}} \frac{dK}{d\Lambda}\frac{1}{\Lambda^2}d\Lambda$ can be calculated as an input before numerically solving the equations. Now the equations can be solved by a number of methods: lattice Boltzmann method [15], Monte-Carlo method [30], discrete ordinates method (DOM) [27,28,35,45]. Recently, a machine learning method is also developed to solve the non-gray BTE [46]. The computational cost of these methods clearly increases with the number of bands. Taking the DOM-finite volume method as an example, the total number of algebraic equations is $N = N_{band} \times N_{cell} \times N_{angle}$. The time complexity of solving linear systems of equations is between $O(N)$ and $O(N^3)$ (depends on the specific algorithm and the sparsity of the coefficient matrix) [47]. The space complexity is $O(N)$. Therefore, adopting fewer bands can lead to both lower time cost and memory cost.

### III. Band discretization scheme and analysis of numerical accuracy

#### a) Previous scheme based on trapezoidal rule

Previous studies have adopted many band discretization schemes. Most of them are based on frequency-dependent expression (Eq. (1)) and only consider the properties in one crystalline orientation [26,27,35], which neglects the complexity of the Brillouin zone. Only those studies based on MFP-dependent expression can incorporate properties in all crystalline orientations [44,48]. Therefore, we analyze the numerical accuracy of their scheme, i.e. trapezoidal rule. In this scheme, several logarithmically equally spaced MFPs are sampled. The phonon temperature for the MFP (not sampled) between two neighboring MFPs (sampled) is linearly interpolated according to the phonon temperature of the two MFPs. As such, when sampling $N$ phonon bands, the MFP $L_b$ and the weight $w_b$ in Eq. (9) is expressed as:

$$L_b = L_{min} \left( \frac{L_{max}}{L_{min}} \right)^{\frac{b}{N-1}}$$

$$w_b = \left( \int_{L_{b-1}}^{L_b} \frac{dK}{dL} \frac{1}{L^m} \frac{L - L_{b-1}}{L_b - L_{b-1}} dL + \int_{L_b}^{L_{b+1}} \frac{dK}{dL} \frac{1}{L^m} \frac{L_{b+1} - L}{L_{b+1} - L_b} dL \right) / \frac{dK}{dL}(L_b) \frac{1}{L_b^m}, m = 1,2 \quad (11)$$

The error induced by the band discretization is due to the difference between the integration and summation in Eq. (9). Therefore, we apply this scheme to Eq. (9) and estimate the error, taking the MFP accumulated thermal conductivity of silicon (Fig. 1 (a)) as an example. More materials will be discussed in Sec. V. Since the real phonon temperature $T$ is not known but needs to be solved for specific problem, we assume that $T^0$ is known and solve Eq. (4) analytically. The phonon temperature $T$ can be expressed in terms of $T^0$:

$$T = T_B e^{-\frac{|r-r_B|}{L}} + \int_0^{r-r_B} \frac{T^0(r'+r_B)}{L} e^{-\frac{|r'|}{L}} dr',\quad (12)$$

where $T_B$ represent the phonon temperature in the boundary. The $r_B$ is the position of the boundary point, and the direction of $r - r_B$ is along the velocity direction $s$. If we assume that $T^0$ is polynomials: $T^0 = a + b|\mathbf{r}'| + c|\mathbf{r}'|^2 + d|\mathbf{r}'|^3$, as we discovered previously [43,49], Eq. (12) is then expressed as:

$$T = T_B e^{-\frac{|\mathbf{r}-\mathbf{r}_B|}{L}} + a\int_0^{\mathbf{r}-\mathbf{r}_B}\frac{1}{L}e^{-\frac{|\mathbf{r}'|}{L}}d\mathbf{r}' + b\int_0^{\mathbf{r}-\mathbf{r}_B}\frac{|\mathbf{r}'|}{L}e^{-\frac{|\mathbf{r}'|}{L}}d\mathbf{r}' + c\int_0^{\mathbf{r}-\mathbf{r}_B}\frac{|\mathbf{r}'|^2}{L}e^{-\frac{|\mathbf{r}'|}{L}}d\mathbf{r}' + d\int_0^{\mathbf{r}-\mathbf{r}_B}\frac{|\mathbf{r}'|^3}{L}e^{-\frac{|\mathbf{r}'|}{L}}d\mathbf{r}'. \quad (13)$$

All terms are related to the MFP only when $L/|\mathbf{r}-\mathbf{r}_B|$ is between 0.01 and 100 (4 orders of magnitude). We plug each term of phonon temperature and MFP accumulated thermal conductivity of silicon into the integrations in Eq. (9) and adopt the trapezoidal rule. The maximum error of all terms with respect to the number of bands is presented in Fig. 1 (b). Three values of $|\mathbf{r}-\mathbf{r}_B|$ including 10 nm, 100 nm, and 1000 nm are checked, i.e., $L/|\mathbf{r}-\mathbf{r}_B|$ for silicon is in [0.1, 10000], [0.01, 1000], and [0.001, 100] respectively (cover different parts of the phonon temperature). As shown in Fig. 1 (b), 60 bands are needed to ensure the error to be smaller than 1 % for all cases. However, previous study showed that adopting tens of bands can lead to unaffordable computational cost for real three dimensional problems [26]. To improve the efficiency of numerical solvers, we propose a new band discretization scheme in the next subsection.

### b) Our scheme based on Gauss-Legendre quadrature

The Gauss-Legendre quadrature is widely used to for complicated numerical integration [35,50], because it integrates all degree $2n-1$ polynomials exactly when using $n$ sample points. There are two important components in Eq. (9): MFP accumulated thermal conductivity $K$ and phonon temperature $T$. Since the MFP accumulated thermal conductivity is known before solving, we analyze the MFP accumulated thermal conductivity $K$ and then decide how to apply the Gauss-Legendre quadrature. The Gauss-Legendre quadrature is better suited for simple polynomials or exponential functions, and thus the MFP accumulated thermal conductivity $K$ is better expressed as simple polynomials or exponential functions. We plot the MFP accumulated thermal conductivity $K$ for several typical semiconductors in Fig. 1 (a). The data for gallium arsenide (GaAs), diamond, and lead telluride (PbTe) are taken from Ref. [51–53]. The data for silicon (Si) at 300 K is obtained from our own first-principles calculations [37,38]. 70×70×70 q-points are used to sample the Brillouin zone. As shown in Fig. 1 (a), the MFP accumulated thermal conductivity exhibits a similar shape for these materials [53]. It is a convex function for small MFP and a concave function for large MFP in the logarithmic coordinate. An inflection point exists in the curve. Since one simple polynomial or exponential function cannot fit the whole curve, we can use two polynomial or exponential functions to fit the curve in the logarithmic coordinate before the inflection point and the curve after the inflection point respectively. The fitting for silicon is shown in Fig. 1 (c). The R-square of the fitting is above 0.99. Therefore, we first divide the MFP domain into two subdomains, i.e., the subdomain before the inflection point ($L_{infection}$ = 63 nm), and the subdomain after the inflection point, as shown in Fig. 1 (c), and then apply the Gauss-Legendre quadrature in each subdomain in the logarithmic coordinate.. In this scheme, the MFP $L_b$ and the weight $w_b$ in the Eq. (9) is expressed as

$$L_b = 10^{\log\left(\frac{L_{infection}}{L_{min}}\right)x_b/2 + \log(L_{infection}L_{min})/2}, w_b = y_b \log\left(\frac{L_{infection}}{L_{min}}\right)/2, L_b \in [L_{min}, L_{infection}]$$
$$L_b = 10^{\log\left(\frac{L_{max}}{L_{infection}}\right)x_b/2 + \log(L_{infection}L_{max})/2}, w_b = y_b \log\left(\frac{L_{max}}{L_{infection}}\right)/2, L_b \in (L_{infection}, L_{max}]$$
(14)

The $x_b$ and $y_b$ is the Gauss point and its weight in the interval [-1,1] based on Gauss-Legendre quadrature rule.

The error with respect to the number of bands is also estimated for our scheme (the same process as that for the trapezoidal rule), as shown in Fig. 1 (d). The error decreases with the increasing number of bands. If the error of 1% is taken as a criterion, 9 bands are sufficient. The convergent number of our scheme is much smaller than the previous trapezoidal scheme.

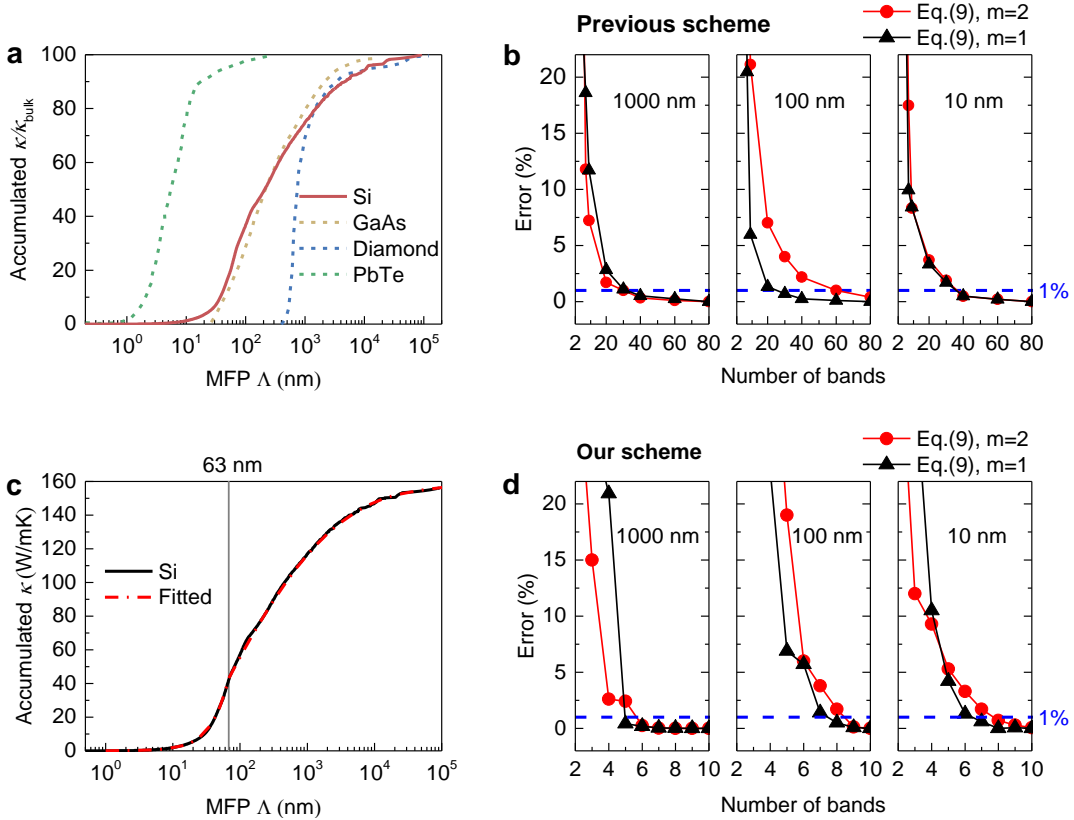

Fig. 1. (a) MFP accumulated thermal conductivity for gallium arsenide (GaAs) [51], diamond [52], lead telluride (PbTe) [53], and silicon (Si). (b) The maximum error among all terms in Eq. (13) of adopting the previous scheme for the integrations in Eq. (9). (c) Fit the MFP accumulated thermal conductivity of silicon. (d) The maximum error among all terms in Eq. (13) of adopting our scheme for the integrations in Eq. (9).

IV. **Numerical verification**

In this section, we numerically solve two real problems to examine the accuracy and efficiency of above schemes and verify the correctness of above estimations, taking silicon as the prototype. Two most important results, thermal conductivity and temperature distribution are checked. For the numerical solver, we select the control angle discrete ordinates method [26] and a standard finite volume method [27] to deal with angular dimensions and spatial dimensions respectively. The first-order up-winding scheme [26,27] is used in the finite volume method. We also adopt the hybrid BTE-Fourier method to speed up the solver [27]. In this method, the phonon bands are divided into two categories according to a cutoff Knudsen number $Kn_{cutoff} = L_{cutoff}/L$, where $L$ is the characteristic length of the problem. The phonon BTE is solved for those bands with $Kn_b = L_b/L > Kn_{cutoff}$ (bands with longer MFP), while the modified Fourier equation is solved for other bands [27]. To solve the algebraic equations, we adopt the sequential method [26,27]. The algebraic equations of each direction and each band are solved by using the Generalized Minimal Residual solver after incomplete LU preconditioning in the PETSc package [54].

a) **Cross-plane heat conduction in the thin film**

We first consider a one-dimensional simulation domain with the cross-plane heat conduction in the thin film, as shown in Fig. 2 (a). We set the thermalizing boundary condition with $T_1$ and $T_2$ for the left and right boundaries, respectively. Three lengths $L$ are tested, including 10 nm, 100 nm, and 1000 nm. The numbers of cells and angles are chosen to be 1000 and 256, respectively after the convergence test. The cutoff Knudsen number $Kn_{cutoff} = L_{cutoff}/L$ is set as 0.01 after the convergence test [27].

We first increase the number of bands in the two schemes to find the band-convergent thermal conductivity and lattice temperature distribution (relative error less than 0.1%). The thermal conductivity is defined as: $k = \dfrac{q}{(T_1 - T_2)/L}$. The two methods can obtain the same convergent results, which shows their correctness (the convergent results are presented in Appendix A). Based on the converged result, we test different number of bands using different band discretization scheme, and the errors are shown in Fig. 2 (b) and (c). The error of adopting the previous trapezoidal scheme is much larger than that of adopting our Gauss method with the same number of bands. By using our scheme, 8 bands can reduce the error to 1% for all cases. In contrast, by using the previous scheme, 40 bands are needed to ensure the error to be smaller 1%, respectively. These numbers are smaller than the estimated value in Sec. III. It is also worth noting that the errors in thermal conductivity and temperature have large difference among different lengths. Simply adopting the convergence test on the bulk thermal conductivity may obtain an underestimated convergent number of bands [26,28,45].

To examine the efficiency, we summarize the time cost for convergent cases. The simulation is performed on a single core on Intel Xeon Scalable Cascade Lake 6248. The CPU time cost is shown in Fig. 2 (d). The time cost of the previous trapezoidal scheme is between 5 times and 10 times using our Gauss method. The time complexity is equal to or above O($N_{band}$), which is consistent with our explanation in Sec. II (c). Since the memory cost is nearly proportional to the number of bands, the memory cost is also reduced accordingly.

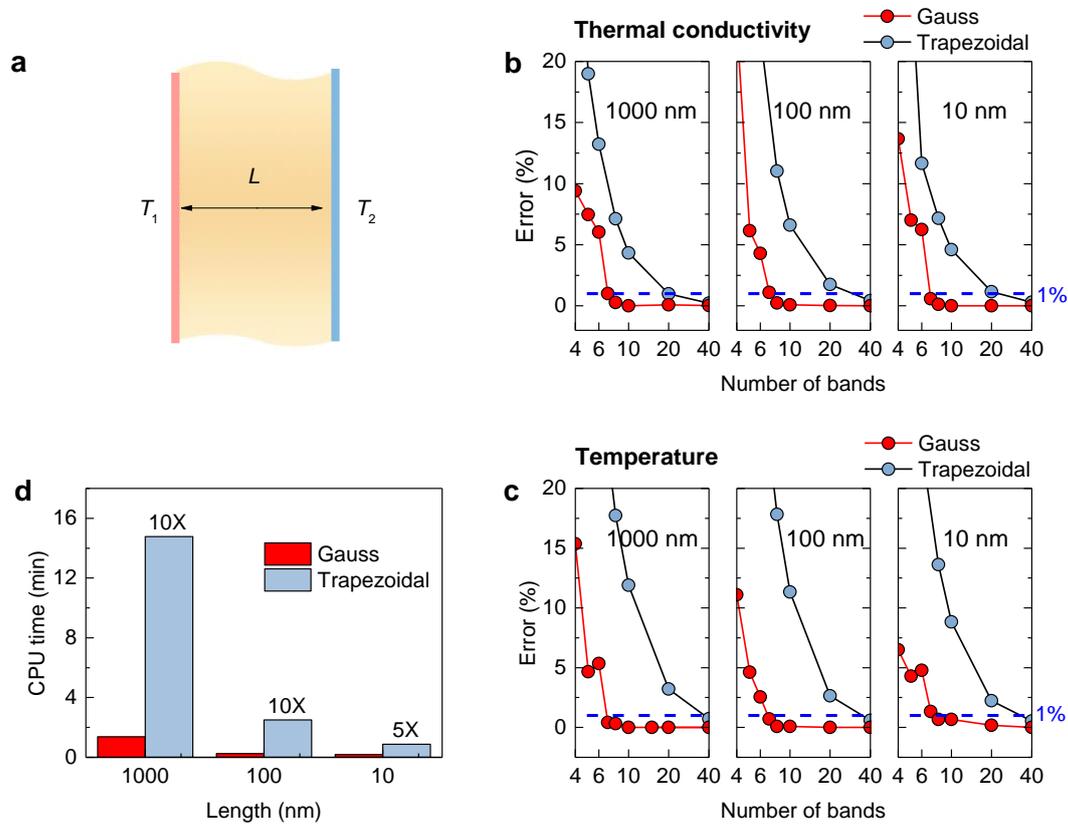

Fig. 2. (a) The schematic of cross-plane heat conduction in the thin film. (b) Relative error of thermal conductivity for cross-plane heat conduction in silicon thin films. (c) Relative error of temperature for cross-plane heat conduction in silicon thin films. (d) CPU time cost of convergent cases of cross-plane heat conduction.

**b) Heat conduction in nano-porous media**

Next we consider a two-dimensional simulation domain with periodic nanopores, as shown in the Fig. 3 (a). The length and the width of the domain are both equal to $L$, while the diameter of the pore is $L/2$. We set the thermalizing boundary condition with $T_1$ and $T_2$ for the left and right boundaries, respectively. The top and bottom boundaries are both set as specularly reflecting boundaries. The boundary of the pore is set as a diffusely reflecting boundary. Three lengths $L$ are tested including 10 nm, 100 nm, and 1000 nm. The numbers of cells and angles are chosen to be 2488 and 256, respectively after the convergence test. The cutoff Knudsen number $Kn_{cutoff} = L_{cutoff}/L$ is set as 0.01, after the convergence test.

We first increase the number of bands in the two schemes to find the band-convergent thermal conductivity and lattice temperature distribution (relative error less than 0.1%). The thermal conductivity is defined as: $k = \dfrac{q_x}{(T_1 - T_2)/L}$. The two schemes yield the same convergent results (the convergent results are presented in Appendix A). Again, we test different number of bands using different band discretization scheme, and the results are shown in Fig. 3 (b) and (c). The error of adopting the previous trapezoidal scheme is generally larger than that of adopting our Gauss method with the same number of bands. By using our scheme, 9 bands can reduce the error to less than 1% for all cases. In contrast, by using the previous scheme, 60 bands are needed to ensure the error to be smaller 1%. These numbers are close to the estimation in Sec. III, which verifies the correctness of the estimation.

To examine the efficiency, we summarize the time cost of convergent cases in Fig. 3 (d). The time cost of the previous trapezoidal scheme is between 11 times and 16 times using our Gauss method.

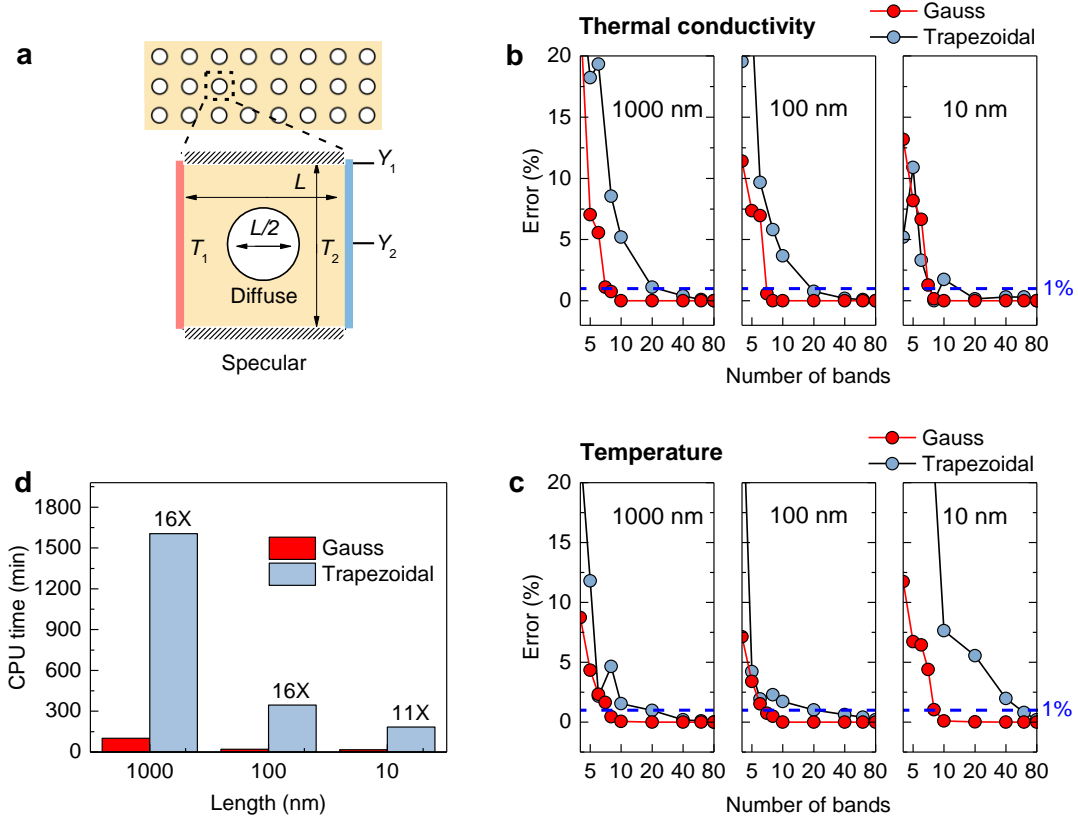

Fig. 3. (a) The schematic of heat conduction in nano-porous medias. (b) Relative error of thermal conductivity for cross-plane heat conduction in silicon nano-porous medias. (c) Relative error of temperature for cross-plane heat conduction in silicon nano-porous medias. (d) CPU time cost of convergent cases of heat conduction in silicon nano-porous medias.

## V. Band discretization for other materials

In the above discussion, we find that our scheme performs much better than the previous scheme for silicon and only requires 9 bands to yield the error less 1%. To check the universality of our scheme, we estimate the error for the all materials presented in Fig. 1 (a) by adopting the process used in Sec. III. Different values of $|\mathbf{r} - \mathbf{r}_B|$ are scanned and the maximum error is shown in Fig 4. For these materials, the error of our scheme (Gauss) is smaller than the error of the previous scheme (trapezoidal) with the same number of bands. For different materials, the error of the Gauss-Legendre scheme does not change much. The reason is that the accuracy of the Gauss-Legendre scheme relates to the shape of the MFP accumulated thermal conductivity. Different materials have the similar shape of the MFP accumulated thermal conductivity. We find that for all four materials we studied, 8-9 Gauss-Legendre bands can yield an error less than 1%, as shown in Table 1. Since the shapes of the MFP accumulated thermal conductivity for many materials are very similar [37,53], we expect that our scheme can be applicable to most other materials, and less than 10 bands can ensure the accurate (error < 1%) solutions of non-gray phonon BTE.

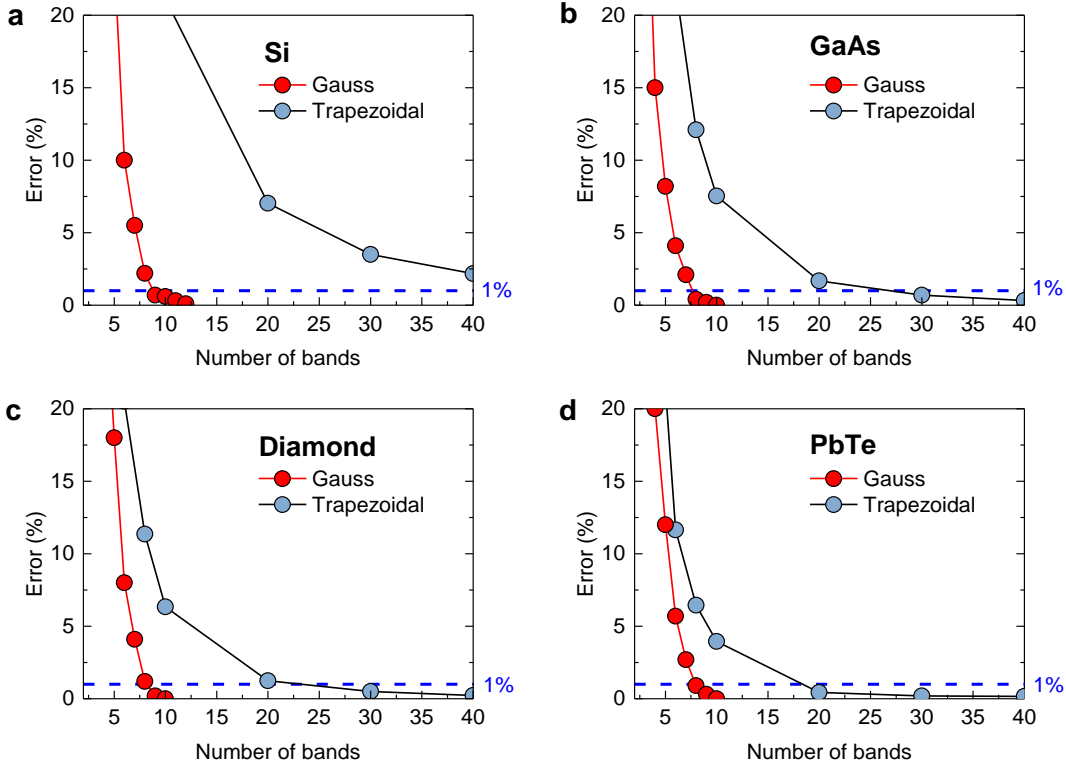

Fig. 4. The error of adopting the our scheme (Gauss) and the previous scheme (Trapezoidal) for the integrations in Eq. (9) for (a) Si, (b) GaAs, (c) diamond, (d) PbTe. The error is the maximum error among all terms in Eq. (13) and different values of $|\mathbf{r} - \mathbf{r}_B|$.

Table 1. Convergent number of bands for various materials

| Materials | Si | GaAs | Diamond | PbTe |
|---|---|---|---|---|
| Convergent number of bands | 9 | 8 | 8 | 8 |

## VI. Summary and conclusions

In this study, we systematically investigate the band discretization of real phonon properties. Based on an analysis of the numerical accuracy and efficiency of band discretization schemes, we propose a scheme to ensure the accuracy with a reduced number of bands compared with the previous scheme. This scheme is to divide the MFP domain into two subdomains based on the inflection point of the MFP accumulated thermal conductivity and perform the Gauss-Legendre quadrature on each subdomain in the logarithmic coordinate. We estimate the error of this scheme analytically and then verify our estimation by numerically solving cross-plane heat conduction and heat conduction in the nano-porous media. The estimation process provides a fast tool for testing convergent number of bands without numerical testing. We find that when using only 9 bands for silicon, the error would be less than 1%. Compared with the reference method, which requires 60 bands, our scheme reduces the time and memory consumption of numerical solvers by nearly one order of magnitude. The performance of our scheme is checked for other materials, including GaAs, Diamond and PbTe. We find that the proposed scheme can be applicable to other materials and in general using less than 10 bands allows an accurate solution (error < 1%) of non-gray phonon BTE. The properties of the convergent Gauss-Legendre bands in this study are presented in the Appendix B, which can also be used for other BTE solvers. The phonon properties for more materials will be continuously updated on our website: https://sites.ji.sjtu.edu.cn/hua-bao/.

**Acknowledgments**

Y.H. and H.B. acknowledge the support by the National Natural Science Foundation of China (Grant No. 51676121). We would also like to thank Ao Wang from Shanghai Jiao Tong University for providing MFP accumulated thermal conductivity of silicon. Numerical simulations were performed on the π 2.0 cluster supported by the Center for High Performance Computing at Shanghai Jiao Tong University.

**Appendix A. Convergent thermal conductivity and temperature distribution**

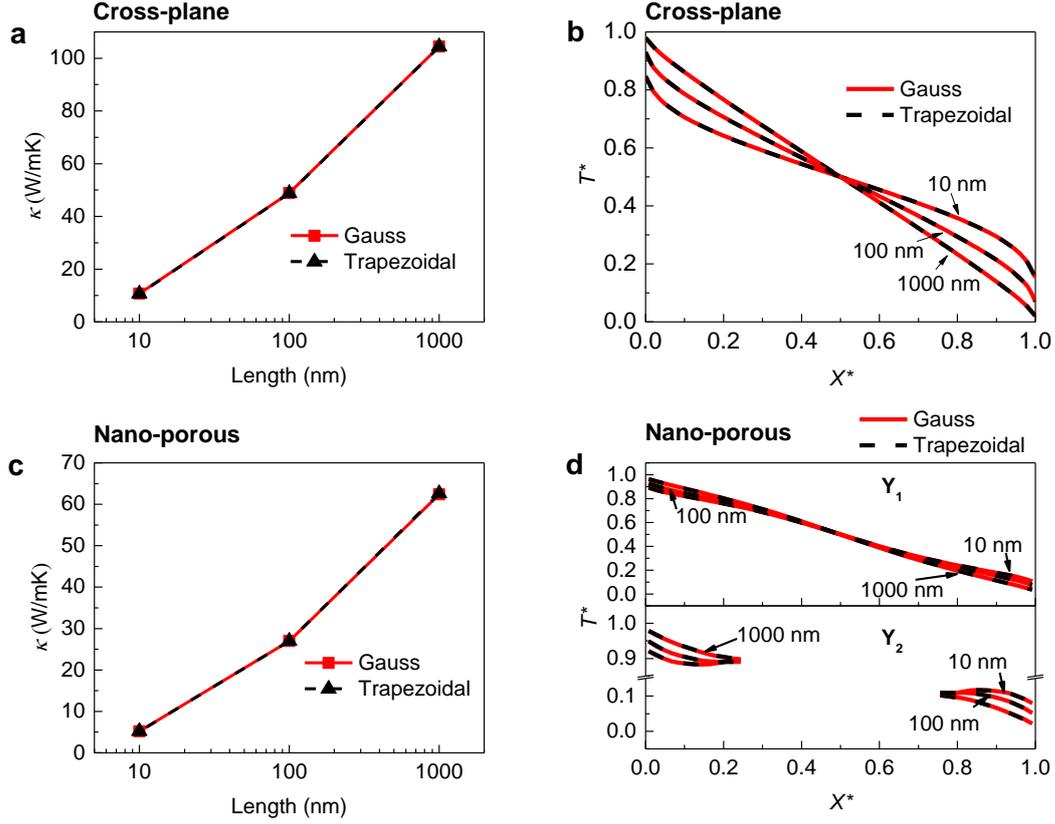

Fig. 5. (a) Convergent thermal conductivity with respect to band discretization for cross-plane heat conduction in silicon thin films. (b) Convergent temperature distribution with respect to band discretization for cross-plane heat conduction in silicon thin films. (c) Convergent thermal conductivity with respect to band discretization for heat conduction in silicon nano-porous medias. (d) Convergent temperature distribution with respect to band discretization for heat conduction in silicon nano-porous medias. The positions of $Y_1$ and $Y_2$ are shown in Fig. 3 (a).

**Appendix B. Phonon band properties**

By applying our band discretization scheme in the non-gray BTE, the non-gray BTE is expressed as:

$$\mathbf{s} \cdot \nabla T_{\mathsf{L}_b,\mathbf{s}} = -\frac{T_{\mathsf{L}_b,\mathbf{s}} - T^0}{\mathsf{L}_b},$$

$$T^0 = \frac{\dfrac{1}{4\rho}\int \sum_{\mathsf{L}_b} \dfrac{W_b}{\mathsf{L}_b} T_{\mathsf{L}_b,\mathbf{s}}\, d\mathsf{W}}{g}, \qquad (15)$$

$$\mathbf{q} = \int \mathbf{s} \sum_{\mathsf{L}_b} W_b T_{\mathsf{L}_b}\, d\mathsf{W},$$

where $W_b = \frac{dK}{dL}(L_b)\frac{1}{L_b}$, $g = \sum_{L_b} \frac{W_b}{L_b}$. These equations can be solved by DOM based methods, Monte Carlo method, lattice Boltzmann method, and so on. The convergent band properties $W_b$ and $L_b$ for the four materials we studied are presented in Table 2.

Table 2. Convergent band properties for various materials

| Silicon | | GaAs | | Diamond | | PbTe | |
|---|---|---|---|---|---|---|---|
| $L_{infection}$ | | $L_{infection}$ | | $L_{infection}$ | | $L_{infection}$ | |
| 6.30e-8 m | | 4.27e-7 m | | 4.02e-7 m | | 9.12e-9 m | |
| $L_b$ (m) | $W_b$ | $L_b$ (m) | $W_b$ | $L_b$ (m) | $W_b$ | $L_b$ (m) | $W_b$ |
| 1.219e-09 | 1.052e+07 | 3.151e-08 | 4.887e+08 | 4.481e-07 | 4.804e+10 | 8.710e-10 | 1.779e+08 |
| 2.646e-09 | 3.592e+07 | 6.526e-08 | 5.605e+08 | 6.027e-07 | 4.117e+11 | 1.371e-09 | 2.014e+08 |
| 8.232e-09 | 9.202e+07 | 1.687e-07 | 2.569e+08 | 8.002e-07 | 2.739e+11 | 2.663e-09 | 6.585e+08 |
| 2.561e-08 | 1.668e+08 | 3.495e-07 | 5.277e+07 | 1.615e-06 | 6.023e+10 | 5.172e-09 | 3.232e+08 |
| 5.561e-08 | 1.395e+08 | 5.733e-07 | 3.850e+07 | 4.826e-06 | 1.056e+10 | 8.140e-09 | 1.148e+08 |
| 1.131e-07 | 1.065e+08 | 1.773e-06 | 1.713e+07 | 1.661e-05 | 1.015e+09 | 1.349e-08 | 9.753e+07 |
| 7.735e-07 | 7.017e+07 | 7.738e-06 | 6.717e+05 | 4.962e-05 | 1.353e+08 | 5.152e-08 | 8.587e+06 |
| 9.504e-06 | 5.323e+06 | 2.393e-05 | 1.076e+05 | 1.002e-04 | 5.509e+07 | 1.967e-07 | 2.567e+05 |
| 6.499e-05 | 1.232e+05 | | | | | | |